\documentclass[fleqn,usenatbib]{mnras}

\usepackage{newtxtext,newtxmath}

\usepackage[T1]{fontenc}
\usepackage{ae,aecompl}
\usepackage{ulem}
\usepackage{graphicx}
\usepackage{color}
\usepackage[dvipsnames]{xcolor}

\usepackage{graphicx}	
\usepackage{amsmath}	
\usepackage{amssymb}	
\usepackage{textcomp}	

\title[Shape Evolution of KBOs]{The Phenomenon of Shape Evolution due to Solar Driven Outgassing for Analogues of Small Kuiper Belt Objects}

\author[Y. Zhao]{
Y. Zhao,$^{1,3}$\thanks{E-mail: zhaoyuhui@pmo.ac.cn}
L. Rezac,$^{2}$
Y. Skorov,$^{2}$
J.-Y. Li$^{4}$
\\
$^{1}$Key Laboratory of Planetary Sciences, Purple Mountain Observatory, Chinese Academy of Sciences, Nanjing 210008, China\\
$^{2}$Max-Planck-Institut f\"{u}r Sonnensystemforschung, Justus-von-Liebig-Weg 3, G\"{o}ttingen 37077, Germany\\
$^{3}$CAS Center for Excellence in Comparative Planetology,
        Chinese Academy of Sciences, China\\
$^{4}$Planetary Science Institute, 1700 E Fort Lowell Road, Tucson, AZ 85719, USA}

\date{Accepted 2019 December 16. Received 2019 December 11; in original form 2019 September 9.}

\pubyear{2019}

\begin{document}
\label{firstpage}
\pagerange{\pageref{firstpage}--\pageref{lastpage}}
\maketitle

\begin{abstract}
One of the key findings of the Rosetta's mission to the Jupiter family comet 67P/Churyumov-Gerasimenko was its peculiar bilobed shape along with the apparent north/south dichotomy in large scale morphology. This has re-ignited scientific discussions on the topic of origin, evolution and age of the nucleus. In this work we set up a general numerical investigation on the role of solar driven activity on the overall shape change. Our goal is to isolate and study the influence of key parameters for solar driven mass loss, and hopefully obtain a classification of the final shapes. We consider five general classes of three-dimensional (3D) objects for various initial conditions of spin-axis and orbital parameters, propagating them on different orbits accounting for solar driven CO ice sublimation. A detailed study of the coupling between sublimation curve and orbital parameters (for CO and H$_{2}$O ices) is also provided. The idealizations used in this study are aimed to remove the ad-hoc assumptions on activity source distribution, composition, and/or chemical inhomogeneities as applied in similar studies focusing on explaining a particular feature or observation. Our numerical experiments show that under no condition a homogeneous nucleus with solar driven outgassing can produce concave morphology on a convex shape. On the other hand, preexisting concavities can hardly be smoothed/removed for the assumed activity. In summary, the coupling between solar distance, eccentricity, spin-axis and its orientation, as well as effects on shadowing and self-heating do combine to induce morphology changes that might not be deducible without numerical simulations. 
\end{abstract}

\begin{keywords}
comets:general, comets:67P/Churyumov-Gerasimenko, Kuiper belt:general
\end{keywords}



\section{Introduction}
Jupiter family comets (JFCs) \citep{Weissman:1991} are known to contain a large fraction of volatile \citep{Mumma:2011} and super-volatile ices \citep{Reach:2013}, which, combined with dynamics of evolving solar system, indicates their likely origin to be in the cold environment of the so called trans-Neptunian region \citep{Duncan:2004,Lowry:2008} . A precise determination of their formation region is difficult to pinpoint from elemental composition point of view \citep{Brownlee:2006,Burchell:2009}, as well as whether the accretion process follows the hierarchical accretion \citep{Davidsson:2016}, or gravitational instability scenario \citep{Lorek:2016}. A significant influence also comes from the subsequent dynamical history of the solar system \citep{Morbidelli:2015}. Additional difficulties arise when aiming at disentangling their original size distribution \citep{Meech:2004,Nesvorny:2018}, dynamical \citep{Levison:1997,Emelyenenko:2013} as well as collisional \citep{Davis:1997,Morbidelli:2015,Jutzi:2017a} evolution before becoming today's observed JFCs.

Therefore, it is not possible in practice to quantify which specific key characteristics of JFCs, such as shape, morphology, bulk density, porosity, observed heterogeneity of outgassing, and etc, retain (and to what degree) their ``primordial'' origin \citep{Massironi:2015,Mousis:2017,Schwartz:2018,Nesvorny:2018}. It is then still necessary to carry out numerical investigations to understand the evolution of the main cometary characteristics in order to be able to piece together what these primitive bodies can tell us about conditions of the protoplanetary disk.

In this connection, one of the key findings of the Rosetta mission \citep{Glassmeier:2007,Taylor:2017} to the JFC 67P/Churyumov-Gerasimenko was the bilobed shape of the nucleus with pronounced morphological features, such as pits, large cliffs, and smooth planes \citep{Sierks:2015a,Thomas:2015,Vincent:2015}. The formation mechanism of the peculiar bilobed shape as a result of a gentle, sub-catastrophic collision between two fully formed cometesimals was suggested by \citep{Massironi:2015}. They also noted that the structural similarities between the two lobes indicate they experienced similar primordial stratified accretion. In the work of \citet{Hirabayashi:2016}, the bilobed shape is a result of fission due to sublimation torque induced spin up in its early history, with subsequent reconfiguration into its present form. \citet{Jutzi:2017b} proposed a re-accumulation induced formation mechanism following a sub-catastrophic collision of a parent body with a smaller impactor. Subsequently, \citet{Schwartz:2018} showed that the bilobed or elongated comets can be formed in the wake of catastrophic collisional disruptions of larger bodies while maintaining their volatiles and low density throughout the process. They argued that since this process can occur in any epoch of our Solar System's history, there is no need for these objects to be formed primordially.

Another possible scenario responsible for nucleus shape modification is sublimation \citep{Jewitt:2009, Jewitt:2017}. Most recently, simulations done by \citet{Vavilov:2019} present the nucleus shape evolution processes from a sphere to varieties of irregular structures including bilobed ones due to anisotropic mass loss. Their work assumes a nucleus with higher production rates in the center region of the comet such that the bilobed/concaved structure can be produced. They also propagated their object in a simplified model ignoring the shape/mass loss induced spin axis and spin rate changes. 

In this work, we aim to extend and improve on the study of \citet{Vavilov:2019} in the effort to understand the limits of possible solutions of long-term sublimation driven mass loss in more realistic way. We combine spin and orbits states with idealized outgassing model of CO at large heliocentric distance acting over long time scales. Discussion of applicability to sublimation of water and methane ices is also provided.

The paper is structured as follows. In section 2 we provide detailed description of the initial orbital parameters, shape models and their basic characteristics, as well as the spin-orbit propagation model. Section 3 contains description of the simulation results. In section 4, we will summarize the main lessons learned and their implications, as well as model approximations and limitations. Furthermore, we outline the future work to be carried out to form a fully self-consistent phenomenology of shape changes due to illumination driven outgassing.

\section{Numerical modeling}
In this section, we provide details of the numerical model built to investigate the sublimation driven mass loss and its effects on reshaping the nucleus. 

\subsection{Nucleus shapes}
\label{sec:shapes} 
\begin{figure}
\includegraphics[width=\columnwidth]{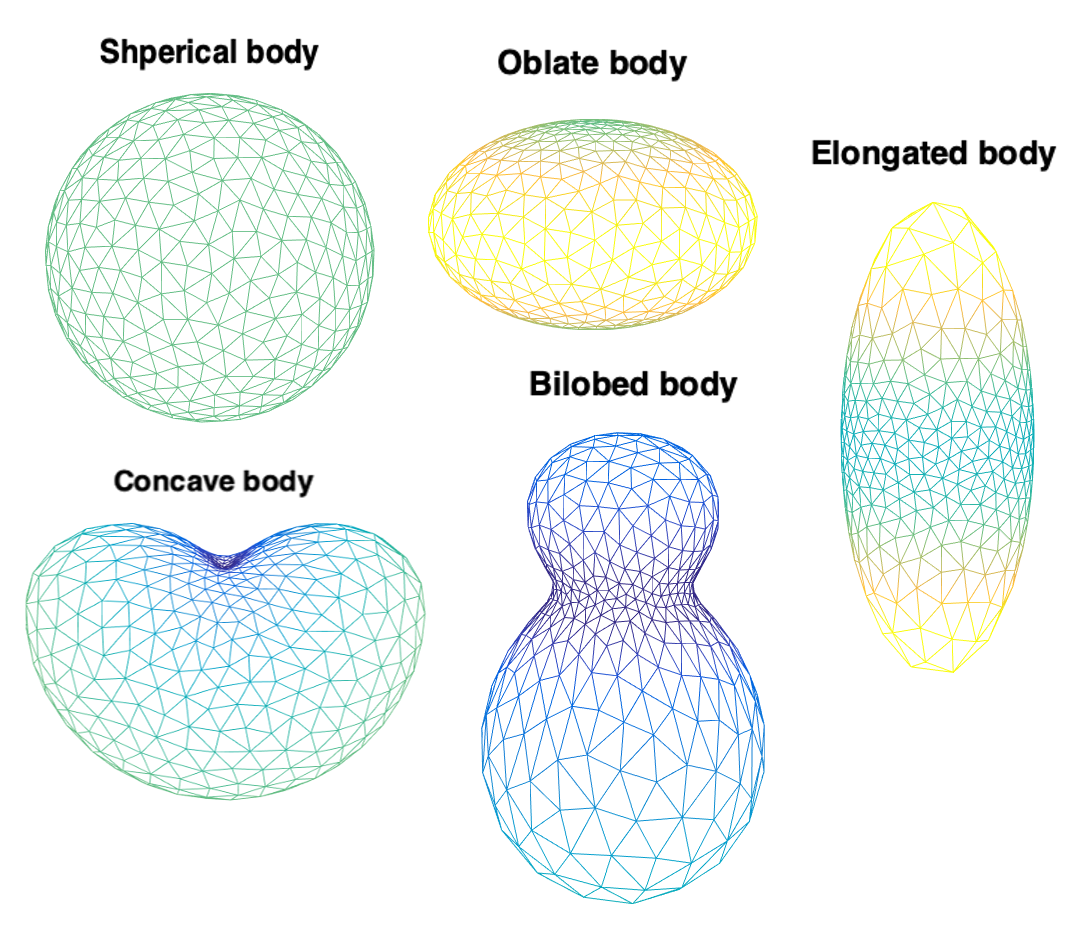}
\caption{Five shape models used as initial bodies in our numerical simulations. The edge color indicates the surface height (body-centric distance, see details in the text). }
\label{fig:simulation_shape.png}
\end{figure}

In this work, the nucleus shapes are built base on 3D triangulation method where each shape is described in terms of surface vertices and facets. An iterative process is employed to generate 500 of approximately uniformly distributed surface grid points on a sphere, and a Delaunay triangulation method is used to achieve the surface triangulation \citep{Field:1988}. As shown in Fig.~\ref{fig:simulation_shape.png}, we will investigate the evolution of five, what we consider, ``general'' shapes. 1) a spherical shape, 2) an oblate body rotating around the axis of the maximum moment of inertia that is often regarded to be an empirical shape model for solar system small bodies, 3) an elongated shape which could be evolved from 1) and 2) \citep{Vavilov:2019}, 4) a concave shape exhibiting a small initial concavity to study its evolution under different starting conditions, and finally, 5) a bilobed shape where the initial configuration resembles that of a contact binary. In addition, we consider two cases of spin-axis for each body, a) the long axis rotation and b) the short axis rotation. The later four shapes are transformed from the original spherical one, and the body-centric distance is measured relative to the center of the sphere rather than the mass center of the resulted shape.

A tri-axial body (sphere, oblate and elongate shape) can be described by three axes, a, b and c:
\begin{equation}
    \frac{x^2}{a^2}+\frac{y^2}{b^2}+\frac{z^2}{c^2} = 1,
\end{equation}
and, the following formula determines the radial distance of a specific surface grid point:
\begin{equation}
    r = abc/(b^2c^2\cos^2\delta\cos^2\alpha+a^2c^2\cos^2\delta\sin^2\alpha+a^2b^2\sin^2\delta)^{\frac{1}{2}},
    \label{eqn:trizx}
\end{equation}
where $\alpha$ and $\delta$ are longitude and latitude of surface grid point, respectively, in the nucleus body fixed frame with the origin at the center of the original spherical shape. For a sphere, the axes have relationship $a=b=c$, an oblate body corresponds to condition $a=b>c$, and an elongated one has $a>b=c$. The parameters we use for simulation in this work are listed in Table \ref{tbl:abc}.
\begin{table}
\caption{Parameters used to define the tri-axial bodies in our simulations}
\begin{center}
{ \hfill{}
\begin{tabular}{p{4cm} c c c c }\hline\hline\noalign{\vspace{1ex}}
Shape & a (${km}$)& b (${km}$) & c (${km}$) \\
\hline
Sphere & 10 &  10  & 10 \\
Oblate body & 16  &  16 & 10 \\
Bilobed body & 10 & 10 & 16 \\
\hline
\end{tabular}}
\hfill{}
\label{tbl:abc}
\end{center}
\end{table}

For the concave shape, we adopt the expression used by \cite{Gutierrez:2000} for aspherical surface with small perturbations and local topographical structures. The body-centric distance (in km) of a surface point in terms of spherical coordinate $(\alpha, \delta)$ measured relative to the spherical center is defined by
\begin{equation}
    r = 10 *(1.0 - 0.8*\cos^2(\delta)*\cos^2(\alpha/2)).
    \label{eqn:bilobe}
\end{equation}

A similar expression is employed to depict the bilobed shape of a contact binary, the corresponding body-centric distance (in km) of a surface point is determined following the equation
\begin{equation}
r = 
\left\{
             \begin{array}{lr}
             10 *(1.0 - p_{n}*\cos^2(\delta))*(1.0 - p_{s}*\sin\delta), & \delta > 0  \\
             10 *(1.0 - p_{n}*\cos^2(\delta))*(1.0 + p_{l}*\sin\delta). & \delta < 0
             \end{array}
\right.
\end{equation}
Points with $\delta>0$ are located at the surface of the smaller lobe and for $\delta<0$ at the larger lobe. $p_{n}$ defines the thickness of the neck region while $p_{s}$ and $p_{l}$ describe the scales of the smaller and larger lobes respectively. The bilobed shape shown in Fig. \ref{fig:simulation_shape.png} has the values of 0.82, 0.6 and 0.2 for these three parameters and the dimensions of each lobe are about 4~km and 8~km respectively. A shape with the two lobes of equivalent dimensions is also studied in our simulation.

\subsection{Sublimation model}
\begin{figure}
\includegraphics[width=\columnwidth]{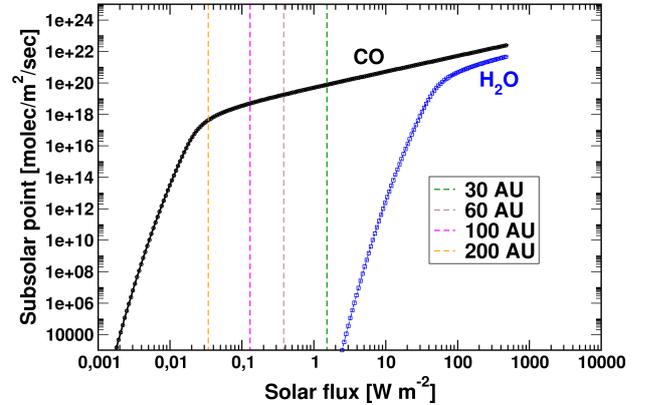}
\caption{Mass loss rate of CO (black) and H$_{2}$O (blue) due to sublimation as a function of solar flux. The dashed vertical lines are provided as an example of incoming flux values at several heliocentric distances as shown in the labels. An important feature to note is the sharp drop in sublimation flux, the ``knee'' in these curves. Although these transitions happen at very different values of incoming fluxes (or heliocentric distances) for water and CO, the coupling of these curves to orbital parameters and the body's orientation states can be generalized as discussed in the text.} 
\label{fig:fluxdist.png}
\end{figure}

In this work, we consider ideal, homogeneous objects composed of pure CO ice, loosely considered to be the analogues of Kuiper Belt objects, as we investigate the upper limits of mass loss in our simulations at heliocentric distances beyond 30~au. The CO (and H$_{2}$O for comparison) sublimation pressure as a function of temperature is taken from \citet{Fray:2009}. For our simulations, we pre-calculate a table of mass loss rate (kgm$^{-2}$s$^{-1}$) as a function of incoming solar flux considering the surface energy equilibrium condition as
\begin{equation}
    E_{sol} = E_{ir}(T) + E_{sub}(T).
    \label{eq:ene1}
\end{equation}
Where E$_{sol}$ is the incoming solar flux, E$_{ir}$(T) represents infrared re-emission flux which we consider with efficiency of 95\% here. E$_{sub}$(T) is energy lost in sublimation of CO ice and depends on the value of latent heat of sublimation which was taken from \citet{Womack:2017}. Both terms on the right hand side are function of (surface) temperature $T$ in our formulation. The resulting mass loss functions are shown in Fig.~\ref{fig:fluxdist.png} for both CO and H$_{2}$O ice, for comparison. The figure also shows lines of constant incoming flux corresponding to several heliocentric distances. Although we consider only CO mass loss in the main text of this paper for morphological evolution of shapes, our conclusions are also applicable for mass loss due to water and methane outgassing. That is, the identified key coupling characteristics between the orbital, spin and mass loss function remain the same but apply at different heliocentric distances (and naturally occur at different time scales). We will discuss this in more detail later. 

\subsection{Shadowing and self-heating}
Since we are interested in generalization of sublimation driven mass loss effects on morphological shape change, our model accounts for the effects of shadowing and self-heating. Naturally, these effects play a role only for non-convex features on the shapes. That is, for each facet, $i$, we calculate the so called ``visibility'' factor, which is a [0,1] flag whether a sun ray can hit this particular facet (at given orientation) or whether it is shadowed by any other facet of the body. If facet, $i$, can receive a direct solar radiation, a cosine of solar zenith angle is multiplied into this factor.

A self-heating effect is relevant for concave feature of the body that may be either pre-existing or developed during the numerical evolution. This effect requires a modified approach to the one presented in Eq.~\ref{eq:ene1} for convex shapes. The energy equilibrium at the surface obtains another term accounting for thermal radiation received at facet, $i$, from all other facets enumerated by $j$ that are visible from $i$. Moving all radiative terms to the left side of the balance, we can write
\begin{equation}
    E_{sol} + E_{df} - E_{ir}(T) = E_{subl}(T).
\end{equation}
For a given facet, $i$, the $E_{df}(i)$ represents diffuse (infrared) flux contribution from all other facets to the facet $i$, and following \citet{ozisik:1985} and \citet{davidsson:2014} can be written as
\begin{equation}
    E_{df}(i) = \sigma \sum_{j \neq i}^{N} F_{ij}T_{j}^{4},
\end{equation}
where the sum goes over all (visible) facets and F$_{ij}$ is the viewing factor between two facets $i$ and $j$, which depends on the cosine of the angles between each facet's normal direction and the ray connecting the two facets, the square of distance between the facets, $d_{ij}$, and the area of facet $j$, $A_{j}$:
\begin{equation}
    F_{ij} = \dfrac{A_{j}\cos{\phi_{ij}}\cos{\phi_{ji}}}{\pi d_{ij}^{2}}.
\label{eqn:selfheating}
\end{equation}
Where $\phi_{ij}$ is the angle between normal direction of facet $i$ and the position vector of facet $j$ relative to facet $i$.

For the simulated objects that feature a concavity and continuously lose mass, hence modify their morphology and shape, the self-heating and shadowing factors have to be often recalculated. These terms are evaluated following the calculation of surface depth change and shape modification due to mass loss for each orbit, as discussed in \ref{sec:rot}. For convex shapes the self-heating terms do not need to be considered, and as will be discussed later, concave features can not develop on an initially convex shape for any of the different simulation conditions.

\subsection{Orbit and body rotation}
\label{sec:rot}
In this work, we will not focus on any specific orbits or discuss dynamic evolution of the objects, instead we aim at demonstrating how orbital parameters could affect the shape modification of the nucleus. As we show in Fig. \ref{fig:fluxdist.png}, to study pure CO driven activities and to better demonstrate the effects, we prefer to use orbits with perihelion and aphelion for orbits on both sides of the 'knee' which is at about 200~au for CO ice, therefore, in most of our simulations, the semi major axis $a$ is set to be 200~au with orbital period $T_{orb}$ about 2800~years. However, orbits with smaller $a$ but larger eccentricity $e$ will also fulfill this condition, which we will discuss in detail in Section \ref{sec:results_ecc}.

\citet{Lacerda:2006} listed rotational periods $T_{rot}$ of more than 10 KBOs (in their Table 10), which range from about 4.7 to 17.7 hours. Here we adopt a reasonable value of 10 hours for $T_{rot}$ in our simulations, this value would not affect the simulation results in a significant way as long as $T_{rot} << T_{orb}$.

\begin{figure}
\includegraphics[width=\columnwidth]{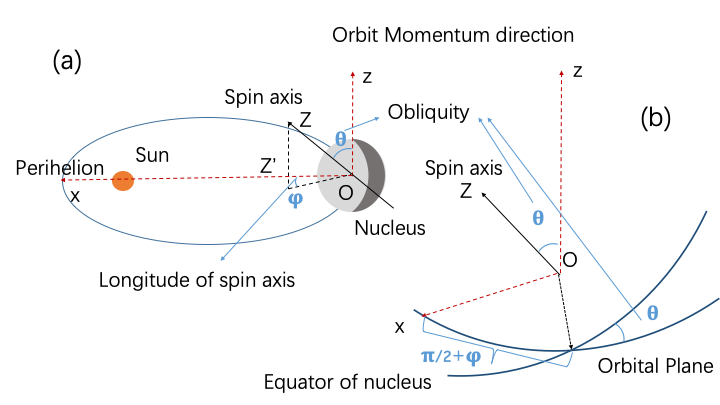}
\caption{Longitude $\varphi$ and obliquity $\theta$ of nucleus orientation in orbital frame. $\vec{Oz}$ and $\vec{Ox}$ point to the orbital momentum direction and perihelion respectively, $\vec{OZ}$ is the rotation momentum direction of the nucleus, and the orientation of the spin axis when we consider principal rotating objects. $\vec{OZ'}$ is the projection of $\vec{OZ}$ on orbital plane, longitude $\varphi$ is the angle between $\vec{OZ'}$ and perihelion direction $\vec{Ox}$ following right hand rule, while obliquity $\theta$ is the angle between orbital angular momentum and rotation angular momentum. Graphic (a) shows these two angles with the nucleus at aphelion, (b) shows the relationship between the orbital plane and the equator of the nucleus, as well as the first two Euler angles for transfer from orbital frame to body fixed frame.}
\label{fig:angles.png}
\end{figure}

To demonstrate the spin states of the nucleus relative to its orbit, we employed the orbital frame as the inertia frame in our simulations. The $z$ axis points to the direction of orbital angular momentum which is perpendicular to the orbital plane and the direction of perihelion is considered as the $x$ direction. The orientation of nucleus' spin axis could then be described by its longitude $\varphi$ and obliquity $\theta$ in the orbital frame. As illustrated in Fig~\ref{fig:angles.png}(a), $\varphi$ is the angle of the rotation momentum's projection on orbital plane relative to the perihelion direction, while $\theta$ is the angle between orbital momentum and rotation momentum. If a 3-1-3 set of Euler angle is used and the rotational angle of the nucleus is $\psi$, we have the following transfer matrix for a vector from orbital frame $\vec{r}$ to its expression in body fixed frame $\vec{R}$ as shown in Fig~$\ref{fig:angles.png}$(b),

\begin{equation}
    \vec{R} = R_z(\psi)R_x(\theta)R_z(90^{\circ}+\varphi)\vec{r},
\end{equation}
where $R_x$ and $R_z$ are the basic rotation matrices about $x$ and $z$ axis respectively.

To investigate total sublimation and mass loss for a certain nucleus shape in one orbit, we divide the orbit into $n_{orb}$ steps with equal eccentric anomaly change $\Delta E$ and rotation period $T_{rot}$ into $n_{rot}$ steps with equal time duration. For anytime in one orbital period, the illumination condition on nucleus surface is defined by the combination of the orbiting stage $n_{orb_i}$ and rotating stage $n_{rot_j}$. 

Suppose for the $i$~th orbital step, the nucleus complete $N_{i}$ rotation periods and in the remainder complete $n_{i}$ rotational steps,
\begin{align}
    N_{i} &= \left\lfloor\frac{T_{orb_i}}{T_{rot}}\right\rfloor \\
    n_{i} &= \left\lfloor\frac{T_{orb_i}-N_{i}*T_{rot}}{T_{rot}/n_{rot}}\right\rfloor,
\end{align}
where $T_{orb_i}$ is the duration of the $i$~th orbiting stage, and could be calculated from mean anomaly change $\Delta M$:
\begin{align}
    T_{orb_i} &= (M_i^e-M_i^i)/\omega_{orb} \\
    M &= E - e\sin{E},
\end{align}
where E and M are eccentric anomaly and mean anomaly, $M_i^e$ and $M_i^i$ are mean anomalies at the end and beginning of the $i$~th orbit step, which could be derived from the relative eccentric anomaly. $\omega_{orb}$ is the mean motion. For each facet $k$, we sum up sublimation for all rotation steps in $i$~th orbital step,
\begin{equation}
    Q_{i}^k = \sum_{j=1}^{n_{rot}}Q_{ij}^k({T_{rot}/n_{rot}})A_kN_{j},
\end{equation}
where $Q_{ij}^k$ is the production rate per area combining the solar distance of the object during the $i$~th orbital step and its orientation status in $j$~th rotation step for facet $k$. $A_k$ is the area of $k$~th facet. $N_j$ indicates how many times the $j$~th rotation step occurs during the $i$~th orbital step,
\begin{equation}
N_{j} = 
\left\{
             \begin{array}{ll}
             N_{j} = N_{i} + 1, & j <= n_{i}  \\
             N_{j} = N_{i}, & j > n_{i} + 1  \\
             N_{j} = N_{i} + \frac{T_{orb_i}-N_{i}*T_{rot}}{T_{rot}/n_{rot}}-n_{i}. & j = n_{i} + 1
             \end{array}
\right.
\end{equation}

We sum up total production rate through all orbit steps to get total mass loss of facet $k$ in one orbit, $Q^k$, and then obtain the relative depth change $D^f_k$ based on nucleus density $\rho$ and facet area $A_k$.
\begin{align}
 Q^k &= \sum_{i=1}^{n_{orb}}Q_{i}^k\\
 D^f_k &= Q^k/(A_k\rho).
\end{align}

To avoid distortion and re-meshing process, we shift the location of surface vertex instead of the facets themselves. Suppose the $l$~th vertex is shared by $m$ surface facets, therefore, the displacement of this vertex $\vec{D}^v_l$ is the combination of depth changes of all facets sharing it, weighted by the depth changes itself,
\begin{equation}
    \vec{D}^v_l = -\frac{\sum_{k=1}^{m}(D^f_k)^2\vec{n}_{k}}{\sum_{k=1}^{m}D^f_k},
\end{equation}
where $\vec{n}_{k}$ is the normal direction of the k~th sharing facet, of which vertex $l$ is one of the three vertices.

Several tests were made to validate the stability and accuracy of the shape modification process. First of all, we checked many different frequencies of shape modification in one orbit to find that the accuracy of performing shape modification once per orbit is sufficient (for our problem); Secondly, we compared the forward and inverse propagation (inverse mass loss) to validate the stability of the method. An elongated shape was evolved for 12 000 orbits forward and then 12 000 orbits backward after which we calculated the difference between the initial and inversely-evolved shape. This difference was less than 0.5\%, as shown in appendix~\ref{sec:appendix_va1}. Additionally, we compared numerical results against analytical model for a sphere, as shown in appendix~\ref{sec:appendix_va2}, which also shows a very good agreement. Therefore, we are confident that our shape modification process is appropriate for investigation of general global shape changes in this work. Nevertheless, for applications studying finer details of shape evolution for a local or complex morphology, a re-meshing procedure may need to be applied.


\section{Results}
In this section, we provide a comprehensive study of the effects of initial conditions of orientation and orbital parameters on the general tendency of reshaping of the five original bodies due to solar driven sublimation. The goal is to isolate and possibly find a generalization or classification a sublimation process may produce. For instance, important questions are whether sublimation always leads to ``smoothing'' of pre-existing concave morphology, can sublimation alone lead to a strong north/south hemispheric dichotomy, is there a  set of conditions for which convex bodies begin a ``neck'' formation, and what kind of role the heliocentric profile of mass loss rate plays for shape modifications?

For brevity and clarity, we cannot show results of all our simulations. Therefore, at first we focus on providing general principles at work which stood out from simulations. Later, we use these findings to demonstrate and explain several interesting examples of original shape modifications for convex and/or bilobed shapes, including self-heating and shadowing effects for different spin axis positions and orientations.

Before we get to concrete examples we would also like to point out that, in our work, the rotation momentum vector points to the north in the body fixed frame by definition. Since most of our simulations are done with the longitude of spin axis $\varphi\in[0^{\circ},90^{\circ}]$, our object's southern hemisphere is the summer hemisphere for most of our simulation cases, unless otherwise noted.

\subsection{Effect of eccentricity and mass loss function}
\label{sec:results_ecc}
No justification is needed to understand that solar driven sublimation will be strongly influenced by the heliocentric distance where the object finds itself during an orbit. An object with zero eccentricity will be on a circular orbit and experience constant solar flux. By virtue of symmetry the summer and winter hemisphere, even in case of strong obliquity will result in symmetric mass loss for such a (homogeneous) body. On the other hand, a highly eccentric orbit will induce a strong summer/winter difference in mass loss, and then, depending on orientation, the pattern of shape change may be substantially different. Despite this general knowledge, it has not been yet settled what values of eccentricity can be considered high for this scenario to occur, and what kinds of north/south (or summer/winter) dichotomies can be generally induced due to sublimation alone.

Another key parameter that plays equal (or sometimes dominant) role is the heliocentric profile of mass loss rate for a given species of ice in considerations. The sublimation equilibrium pressure depends exponentially on the binding energy of molecular matrix (latent heat of sublimation) and the thermal energy of the matrix. Hence, a sharp decrease in this quantity is expected for very low values of temperature.
This phenomenon could be observed in mass loss curves shown in Fig.~\ref{fig:fluxdist.png} for the case of CO curve (shown in log space). First, there is a "linear" (slope approximately equals to one in log space) decrease with decreasing flux, while at the solar flux of about 0.05 Wm$^{-2}$, the CO sublimation rate starts to decrease dramatically and non-linearly with flux. When the eccentricity of an orbit is such that the orbit of a body crosses this ``knee'' on the mass loss rate curve, we generally observe in our simulations the most extreme dichotomies between the north and south hemispheres. It is instructive to think about heliocentric mass loss rate because even for other processes modifying this function, the role of the ``knee'' remains significant.

\begin{figure}
\includegraphics[width=\columnwidth]{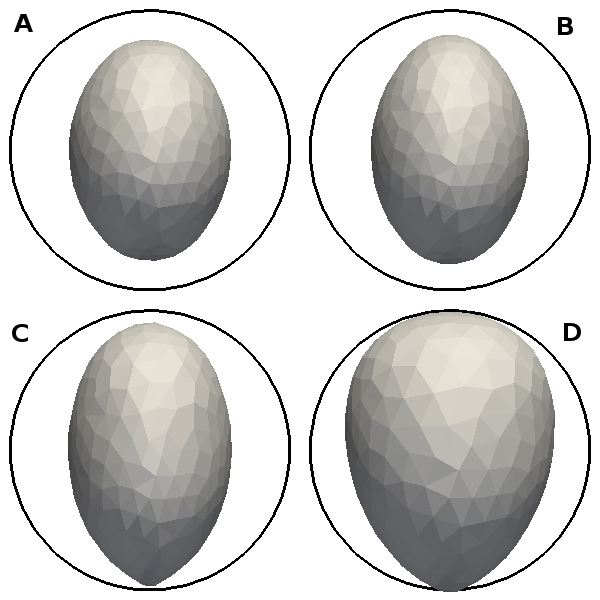}
\caption{An example of dependence of shape change on different configurations of eccentricity and semi-major axis, in the case of an elongated body (shown in a projection along the long axis). Summer hemisphere (south) is at the bottom. In all cases the obliquity, longitude of spin-axis and eccentricity are the same, and set to 30$^{\circ}$, 45$^{\circ}$ and 0.2 respectively. The different panels vary only the semi-major axis to illustrate the effects of moving along the constant eccentricity in Fig.~\ref{fig:fluxratio.png}. In panel A) a=25~au, B) a=75~au, C) a=175~au and D) a=225~au. The black circle is a contour of the original body (same projection) for visual demonstration of the difference with the final shape (see text for discussion).}
\label{fig:elongate4panel.png}
\end{figure}

In order to develop north/south hemispheric differences on a homogeneous original body, it must have a non-zero obliquity. Although a more comprehensive investigation into effects of spin-axis and its orientation will be described in next section, we present a starting point for discussion on the effect of eccentricity in Fig.~\ref{fig:elongate4panel.png}. This particular numerical experiment considers an elongated body shown in Fig.~\ref{fig:elongate4panel.png} in a projection along the long axis (perpendicular to the image plane). A spin axis is parallel to the larger dimension which lies vertically in the plane of the figure. The plotted frame is chosen only so that visible changes are manifested clearly on the shape of the body. The four panels of this figure correspond to four different orbits with the same eccentricity, obliquity and longitude of spin axis [$e=0.2$, $\theta=30^{\circ}$, $\varphi=45^{\circ}$ respectively] (see Fig.~\ref{fig:angles.png} for definition) but different semi major axes [$a= 25, 75, 125$ and $225$~au]. Because the same time frame is shown, bodies at larger heliocentric distances would experience weaker shape modification due to weaker mass loss (e.g. panels C, D). However, the morphological variations among these examples clearly demonstrate different patterns although they all have the same eccentricity, and spin axis orientation. 

Most remarkable is the pronounced north/south morphology that begins to develop for the cases starting at semi-major axis value of about 175~au (panel C), but is already very pronounced for $a=225$~au (panel D). However, for the bodies at $a=25$ or $75$~au (panels A, B) we can discern only a rather symmetric mass loss rate pattern. How can the same eccentricity produce markedly different shape morphology varying only the heliocentric distance? Below we want to formulate a general principle at work, applicable not only to CO but also to other ices (see appendix~\ref{sec:appendix} for H$_{2}$O analysis). By extension, the following argument is also applicable for conditions of presence of a dust layer on the surface (or other processes determining the mass loss rate function). It is worth to have a good understanding of this process even under the ideal conditions in order to interpret the more sophisticated models applied to real observations for the purpose of connecting a shape morphology to its possible evolutionary (outgassing) history.

The eccentricity (for a given obliquity) is not enough to explain the above results of the mass loss pattern. However, according to the second law of Kepler orbit, 
\begin{equation}
    r_p^2\dot{f} = H_{orb}
\end{equation}
as the orbital angular momentum $H_{orb}$ stays constant, the variation rate of true anomaly $\dot{f}$ is inversely proportional to the square of the heliocentric distance $r_p$, hence the time interval of a unit true anomaly $_{\Delta}T_{f}$ has the following relationship with $r_p$ 
\begin{equation}
   _{\Delta}T_{f} \propto r_p^2.
\end{equation}
Meanwhile, the solar flux intensity has an inverse proportion relationship with $r_p^2$.  Since CO sublimation rate is proportional to solar flux inside the non-linear transition zone (the ``knee''), Kepler's 2nd law indicates that the integrated solar energy received at the object, as well as the total mass of sublimated CO from near the sub-solar point, within a unit true anomaly interval remains constant over an orbit. This is independent of the orbital phase of an object or its orbital eccentricity. The main factors causing the north-south dichotomy are the nonlinear dependence of the sublimation rate on solar flux intensity outside of the "knee" and on the sharp decay part in the mass loss function as described in Fig.~\ref{fig:fluxdist.png}, and the relative location of orbit configuration in solar flux space.

\begin{figure}
\includegraphics[width=\columnwidth]{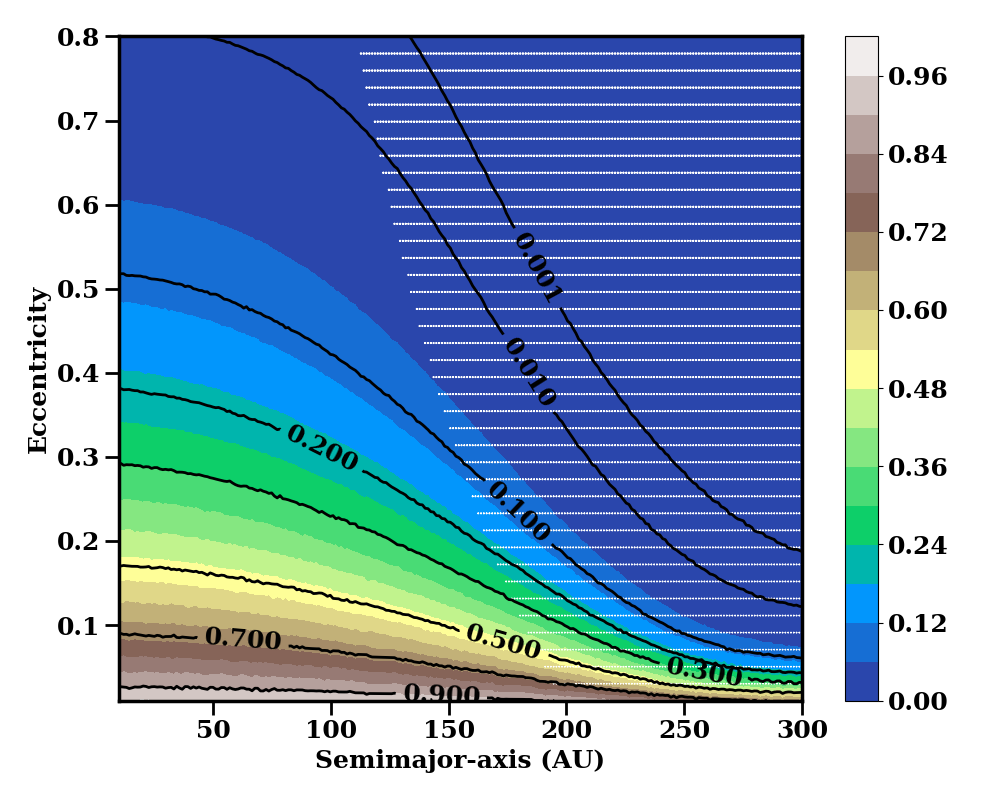}
\caption{A ratio of CO mass loss at apogee to perigee (colorbar) as a function of eccentricity and semi-major axis. This figure illustrates that this ratio can drop significantly for large heliocentric distance and even small eccentricity values, and same can be achieved for small distances but varying eccentricity up to 0.8. White dashed lines overlaid on the plot mark a region where the aphelion of the given orbit lies beyond the ``knee'' of the mass loss rate. See text for discussion.}
\label{fig:fluxratio.png}
\end{figure}

In this case, we found the ratio of aphelion to perihelion mass loss rate, which is related to the curve in Fig.~\ref{fig:fluxdist.png}, to be a key parameter to predict the hemispherical difference. This point is captured in Fig.~\ref{fig:fluxratio.png} showing the CO mass loss rate ratio (aphelion/perihelion) as a function of eccentricity and semi major axis. The white dashed lines overlaid on the plot indicate for which configuration of ($a,e$) the orbit's aphelion would lie at a distance beyond the ``knee'' of the mass loss curve. In this context, results in Fig.~\ref{fig:elongate4panel.png} can be now better understood, since for $e=0.2$ and a$>$~170~au the aphelion of the orbit takes the object beyond the sharp drop in mass loss rate. In this particular setup the winter hemisphere experiences less than 10\% of mass loss compared to the summer.

Nevertheless, there is a number of cases in the ($a,e$) space where the aphelion/perihelion mass loss ratios are very small, but we do not observe strong north/south asymmetries on bodies with such orbits. The reason is twofold. First, in those configurations, the aphelion distances are small enough so that we are still on the flat "linear" part of the mass loss rate curve, the low ratio is a result of the extremely large eccentricity, but the production rate follows the flat "linear" relationship with solar flux, which indicates more uniform mass loss across the orbit. Second, the orbit passes the "knee" near its aphelion, apart from the small fraction of the orbit beyond the "knee", solar flux for most of the orbit locates on the flat "linear" part of the mass loss  rate curve, and hence yields a more average distribution across the orbit. This also indicates the "knee" crossing location on the orbit to be another important factor.

\begin{figure}
\includegraphics[width=\columnwidth]{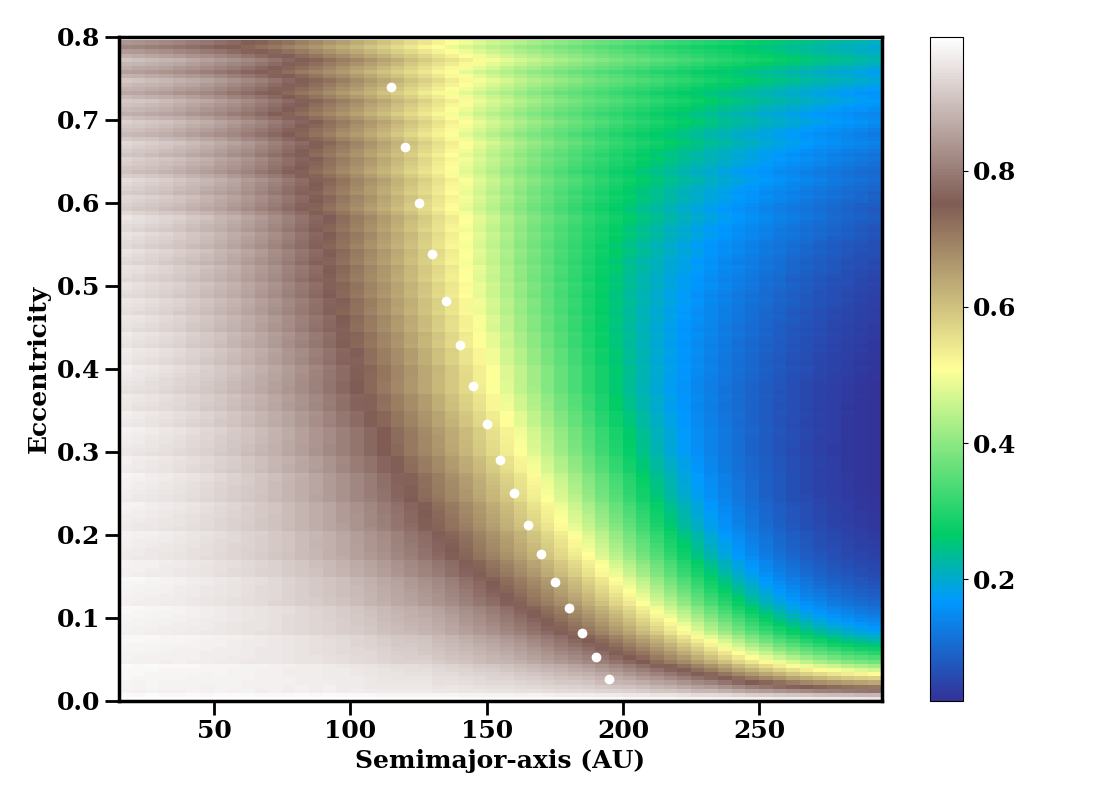}
\caption{A ratio of integrated CO mass loss for aphelion to perihelion segment of the orbit, as a function of eccentricity and semi-major axis. This figure corroborates Fig.~\ref{fig:fluxratio.png}, and provides another view for what configuration ($a,e$) a large hemispheric differences would be occur. The white dotted line indicates a boundary (in geometric sense) beyond which the aphelion cross the ``knee'' of the CO mass loss function. Our simulations show that in this phase space when this ratio is about $<$0.5 objects tend to develop strong north south asymmetry. }
\label{fig:fluxintegral.png}
\end{figure}

A more quantitative way to consider this general principle is to evaluate a ratio of integrals of mass loss at sub solar point for two orbital segments: the semi-orbit with $90^{\circ} < f \le 270^{\circ}$ passing through aphelion and the semi-orbit with $f \le 90^{\circ}$ or $f > 270^{\circ}$ passing through perihelion. This ratio is plotted in Fig. \ref{fig:fluxintegral.png}, which provides a more direct indication of differences between mass loss rates for the two hemispheres. For orbits located on the flat "linear" part of the mass loss rate curve, as the second Kepler's law indicates, total CO mass loss during each true anomaly unit follows the "linear" relationship and is nearly the same everywhere on the orbit. Therefore, the ratio for the two orbital segments of this kind of orbits is close to 1, as shown in Fig. \ref{fig:fluxintegral.png} on the left side of the white dotted line. For orbits with semi major axis larger than heliocentric distance of the "knee", at some certain value of the eccentricity, the ratio reaches the minimum value and starts increasing towards unity as the eccentricity getting larger. This is because that larger $e$ makes the orbit more eccentric, the perihelion distance becomes smaller and the "knee" crossing point on the orbit moves further away from perihelion in true anomaly, a larger portion of the orbit therefore moves into the flat "linear" part of the mass loss function, which makes the mass loss difference between these two semi-orbits getting smaller again.

Based on our numerical experiments, a ratio of integrated mass loss smaller than 0.5 could indicate a strong hemispherical dichotomy due to solar driven sublimation. We consider this as a general conclusion that is also relevant for sublimation driven outgassing due to  H$_{2}$O (see appendix) or CH$_{4}$, although it operates at different heliocentric distances for these molecules. If the icy surface is covered by a uniform dust cover (also an idealization), this conclusion is still valid, because a dust cover will likely just shift the mass loss rate curve with respect to solar flux without changing its shape. In situations where the dust cover affect the location of sharp drop in the mass loss rate curve, the principle is still valid as long as we can consider the body homogeneously active.

\subsection{Effects of obliquity and longitude}
\label{sec:orientation}
With an understanding of how and for which configurations of eccentricity and heliocentric distance an object might experience the largest hemispherical dichotomy, we set out to investigate a sensitivity to other fundamental parameters. In this section, we will consider the effects of obliquity, $\theta$, and longitude of the spin axis, $\varphi$. The graphics in Fig.~\ref{fig:angles.png} define these angles. The longitude of the spin axis can have a strong influence on the heliocentric slopes of total water production curve (eg ., \citet{Marshall:2019}), which is often used for inter-comparisons of activity of different comets. 

\begin{figure*}
\includegraphics[width=\textwidth]{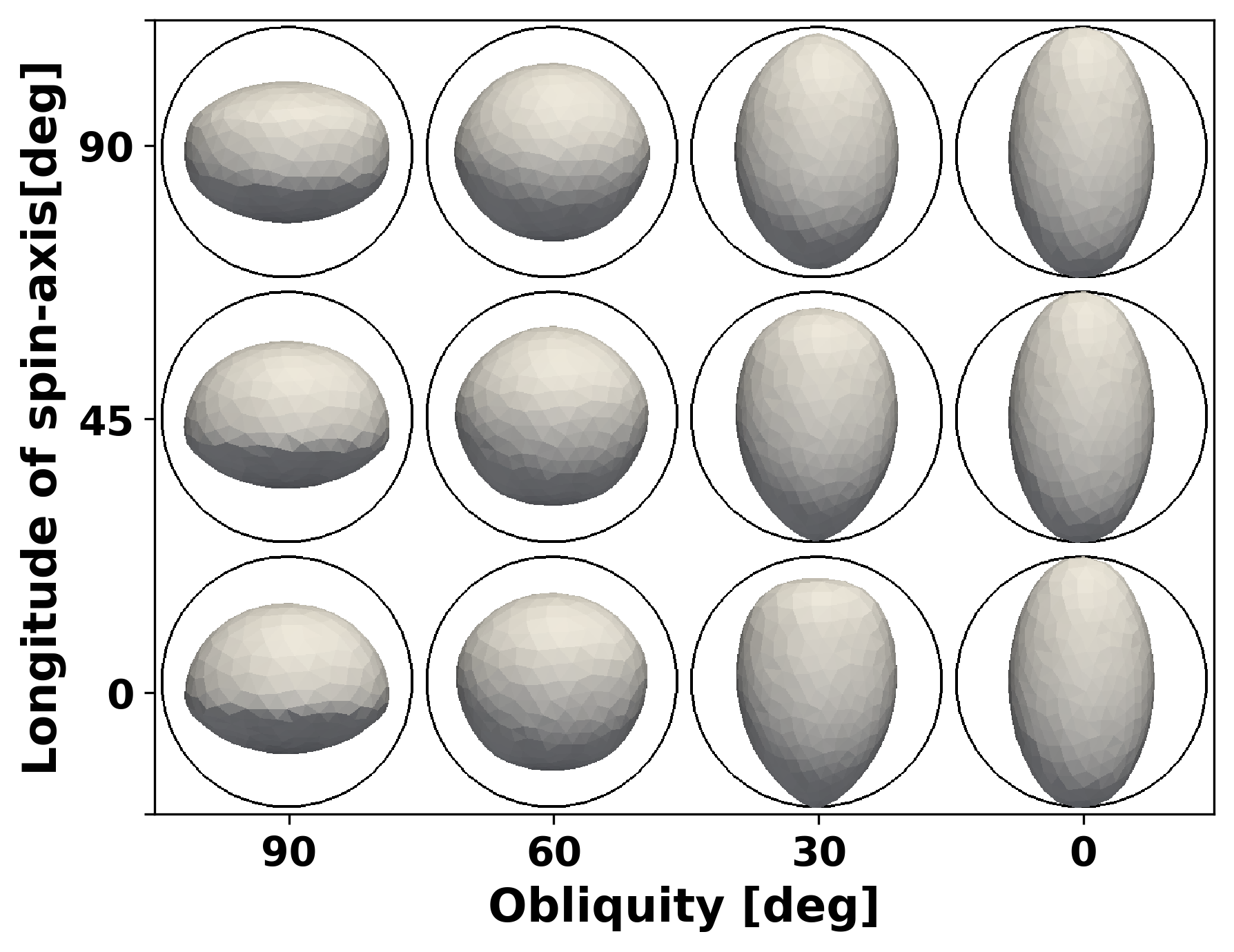}
\caption{The resultant shape changes from initially a spherical object for different configurations of obliquity ($\theta$) and longitude of spin-axis ($\varphi$). The circles mark the size of the original (spherical) body.}
\label{fig:alfdel_sphere.png}
\end{figure*}

Fig.~\ref{fig:alfdel_sphere.png} shows how a homogeneous spherical object would alter its shape due to sublimation for varying values of $\theta$ and $\varphi$. This numerical experiment was run for $a=200$~au and $e=0.3$, which guarantees the potential for a strong hemispherical dichotomy as explained in the previous section. Although we have made simulations for other shapes, the sphere serves as a good starting case to avoid complications which might arise due to more complex shapes, for which a spin-axis definition also begins to play a role (will be discussed subsequently). Similarly as before, the simulation time is considered arbitrary as long as we focus only on the upper limit effects of the shape change due to sublimation. All panels in this figure are for the same simulation time frame.

We may start with the simplest case of $\theta=0$ when the spin axis is vertical to the orbital plane. In this case the sub-solar point is always at the equator which then experiences the largest (and symmetric) mass loss. However, as we increase obliquity to $\theta=30^{\circ}$ and $\varphi=0^{\circ}$, we can already observe the north/south asymmetry in the final shape. The summer hemisphere (bottom in the figure) is eroded much stronger than the winter one as a result of the selected ($a,e$) configuration. As explained we can understand this as much weaker mass loss in the semi-orbit passing through aphelion than it passing through perihelion. Hence, a ``pointy'' looking shape is formed in the southern hemisphere. This pattern is different from $\varphi=90^{\circ}$ (and would be fully reversed for $\varphi=180^{\circ}$, not shown). The most pronounced shape modification in regards to north south asymmetry, and also a strongest dependence on the $\varphi$ is for $\theta=[60^{\circ},90^{\circ}]$. In the case of $\theta=90^{\circ}$ when spin axis lies parallel to the orbital plane, the spherical shape is essentially transformed into an oblate shape, however, depending on the $\varphi$, there are notable hemispherical asymmetries. Generally speaking, for certain configurations of eccentricity, heliocentric distance and a none-zero obliquity, the value of $\varphi$ closer to $0^{\circ}$ or $180^{\circ}$ would produce more pronounced north south asymmetry. Therefore, both obliquity, but also longitude of spin-axis are necessary to know in trying to make a possible link between shape, morphology and evolution of sublimation driven mass loss. 

\subsection{Effects of spin-axis}
\begin{figure*}
\includegraphics[width=\textwidth]{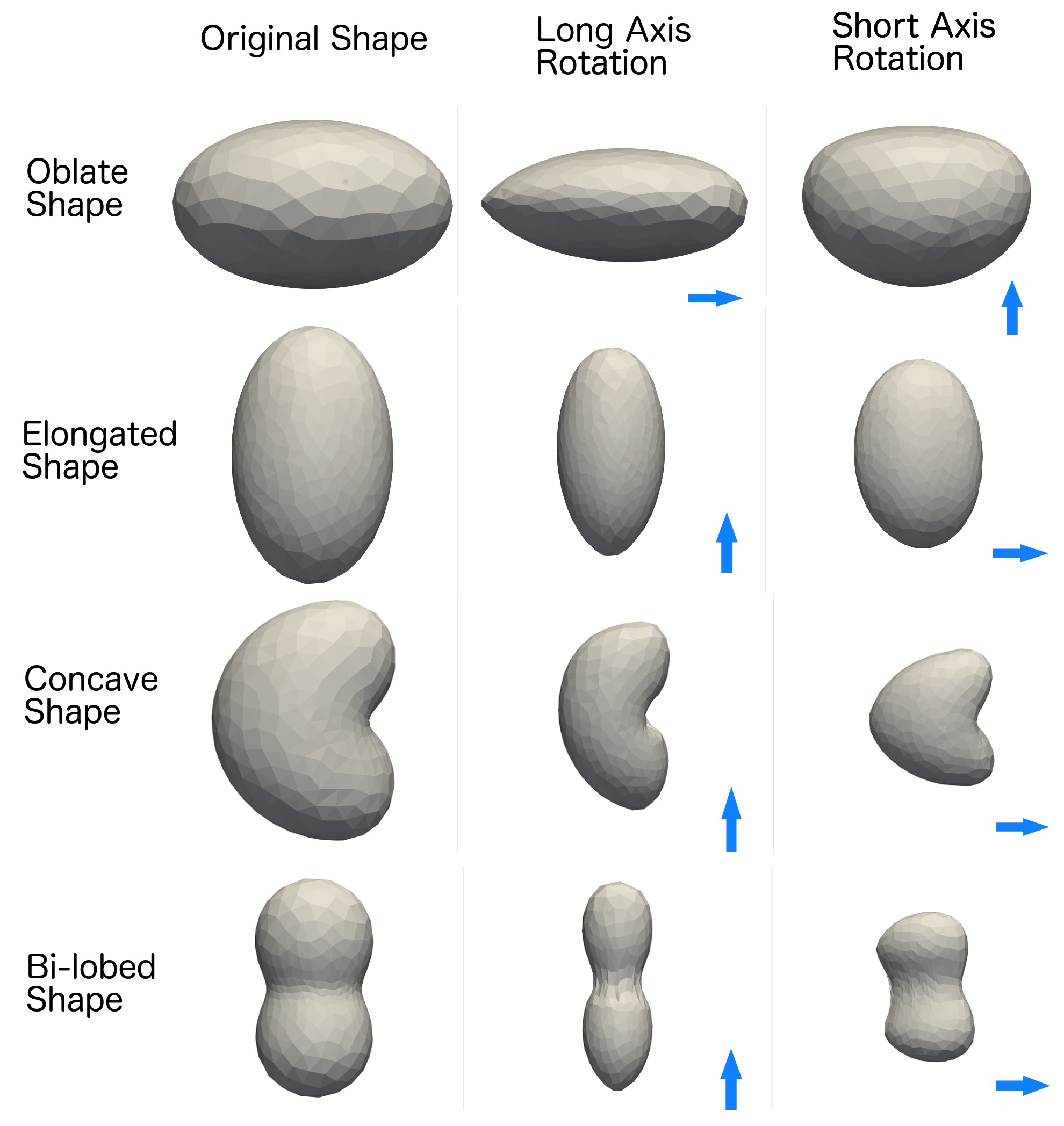}
\caption{Nucleus shape evolution for various original shapes with different spin axis configurations. The blue arrow in the bottom right corner in each panel shows the direction of the spin axis, pointing from the southern pole to the northern pole. The first column shows the original shapes, the second column shows the evolved shapes rotating with their long axis and the third column shows shape evolution with short axis rotation mode. In these simulations we set $a=200$, $e=0.3$, $\theta=30^{\circ}$, and $\varphi=0^{\circ}$, hence, the summer hemisphere is the southern of these bodies.}
\label{fig:shape_axs.png}
\end{figure*}

Other shapes, a bilobed, concave as well as  elongated and oblate share the same character of morphological shape changes as a function of ($\theta$,$\varphi$), as was demonstrated for a sphere. Despite this fact, the altered shapes, especially the objects with concavities, are not straightforwardly predictable. Furthermore, these shapes have also in principle two possibilities of spin axis definition, either to rotate along the shorter or the longer axis. Because it is impractical to present here all the cases of ($\theta$,$\varphi$) for each rotational axis, we demonstrate this effect for a single orbital configuration in Fig.~\ref{fig:shape_axs.png}. This figure shows the resultant shape morphology of the different non-spherical objects for two configurations of different rotation axes. In these simulations we set $a=200$, $e=0.3$, $\theta=30^{\circ}$, and $\varphi=0^{\circ}$. There are four rows in the figure, one for each type of a shape, and three columns with the initial and the evolved object structure for cases of different spin axis. The blue arrow depicts the orientation of the rotation axis. The summer hemisphere is at the southern hemisphere (the base of the arrow). 

Starting with the oblate shape (first row), the possible end results show a strong dependence whether the body is considered as a short or long axis rotator. In general, there is considerable flattening for long axis rotation, also with pronounced north/south dichotomy manifested in the ``sharpenning'' of the southern hemisphere pole. In the case of short axis rotation, although the same principle is at work, the body's erosion leads to a different shape. Because of the rotational symmetry in this case, the ``sharpening'' effect appears to transform the oblate to a spherical looking shape (at the southern hemisphere).

For the elongated shape, we have already described the short axis rotation in connection with Fig.~\ref{fig:elongate4panel.png}. In the case of long axis rotation we also see the north/south asymmetry as the resultant result of sublimation.

The third row catalogues a result for the concave shape. The concavity is located at the equator in the case of long axis rotation (middle column), and for short axis rotation, it is located at the north pole. In the long axis rotation case, there is tendency to make the south pole ``sharper', but more importantly, the morphology of the concave center feature has also changed. It is not flattened due to sublimation, but it has gotten deeper (in relative sense that the entire body is now smaller). The slopes of the depression (concavity) have also been transformed, and they are visibly less symmetric. The short axis rotation induces the largest mass loss at the southern hemisphere, and, due to symmetry, at the side of the body (as expected from the axis orientation). Since the depression is on the north, it experiences only weak erosion. Importantly, we note that although we preformed simulations to cover the entire parameter space of ($a,e,\theta,\varphi$), there was no configuration under which the sublimation process would ``flatten'' the depression as to make it disappear. Although this conclusion pertains only to the pure ice assumption relevant to these simulations, it is an important result. 

Finally, we consider a bilobed shape, with the ``neck'' region parallel to the equator for the long axis rotation, and to the poles for the short axis rotation (right most column). Again, we observe vastly different shapes  produced depending whether rotation along long or short axis is assumed. The initial symmetric depression (boundary where the two spheres join) becomes notably wider at the south pole compared to the north in the case of short axis rotation. In effect, this is the phenomena of north/south asymmetry for bilobed shape. Furthermore, this wider morphology of the ``neck'' as an end result at the summer hemisphere could also be an end results for the concave shape (discussed previously) when the depression would be located at the summer hemisphere. As noted previously, the homogeneous mass loss does not lead to complete eradication of a pre-existing concave morphology.

\subsection{Effects of self-heating and shadowing}
All results of our simulations so far include the effects of shadowing and self-heating, which are relevant for bodies featuring concave morphology. In this section, we explicitly demonstrate importance of such considerations. 

\begin{figure}
\includegraphics[width=\columnwidth]{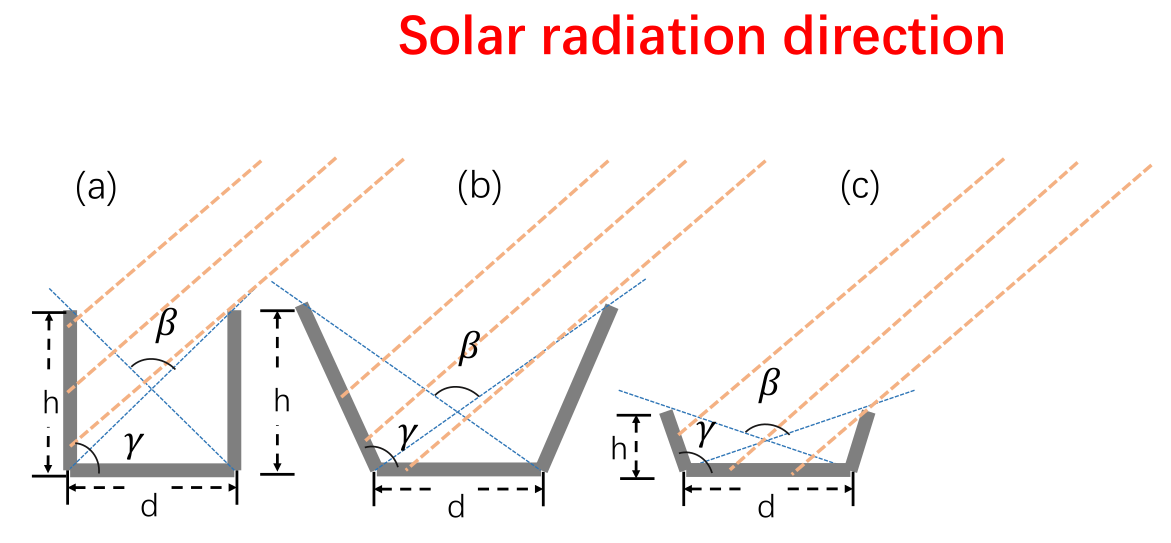}
\caption{A graphic to illustrate the shadowing and self-heating effects work on the erosion process of different concavity structures. $d$ and $h$ are the dimension of the bottom and the height of the concavity respectively, $\gamma$ is the angle for the inclination of the edge relevant to the bottom plane while $\beta$ is the apex angle of the illumination cone as we explained in the text in detail. Three cases are displayed here, graphic a) shows a concave shape of a hole, whose edge is vertical to the bottom with $\gamma = 90^\circ$, graphic b) shows a concave structure which could be used to describe the shape of a crater while graphic c) is the shallower case for b).}
\label{fig:concavity.png}
\end{figure}

The role the shadowing effect plays is strongly dependent on the shape of the concavity. In order to substantiate this argument, we consider a geometrical model with a depth $h$, the bottom dimension $d$ and the inclination of the edge $\gamma$ as shown in Fig.~\ref{fig:concavity.png}. The figure demonstrates how shadowing and self-heating effects work on the erosion process for three different concave structures. It's easy to understand that larger inclination angle $\gamma$ and smaller depth $h$ would result in more solar flux reaching the bottom and the edge of the concavity (case c in Fig.~\ref{fig:concavity.png}). If we define an illumination cone with apex angle of $\beta$ (as shown in the figure), the floor of the concavity may receive solar flux only if the Sun locates in this cone. In such approach, we can express the relationship between the parameters as
\begin{equation}
    \tan(\gamma-\frac{\pi}{2}) + \frac{d}{h} = \tan{\frac{\beta}{2}}.
\end{equation}
The larger the $\beta$ the weaker the shadowing effect (of the surrounding walls), and at the same time the stronger the heating (mass loss) from the bottom (floor). Meanwhile, from the double cosine dependence of the visibility factor (Eq.~\ref{eqn:selfheating}), we can see that $\gamma$ plays a double role. For instance, the smaller the $\gamma$ the less solar flux reaching bottom and also the edge of the concavity (also function of depth) due to the shadowing effects. At the same time, for smaller $\gamma$, the self-heating is expected to be stronger due to larger cosine entering the visibility factor. The combination of these effects will determine the erosion process of the concavity.

For example, consider the case when the bottom of the concavity is illuminated with solar incidence angle of 0$^\circ$, this region gets the maximum of solar flux. In this situation, when $\gamma=90^{\circ}$ (case a in Fig.~\ref{fig:concavity.png}), the edges (walls) of the hole are not illuminated at all and produces no self-heating effects. On the other hand, in the case b), with $\gamma$ larger than $90^{\circ}$, the concavity's wall(s) are under illumination providing additional energy (self-heating) to the the floor of the concave structure which is at the same time also illuminated. This effect was also described by \cite{ivanova:2006}, termed as energy ``concentrator''.

\begin{figure}
\includegraphics[width=\columnwidth]{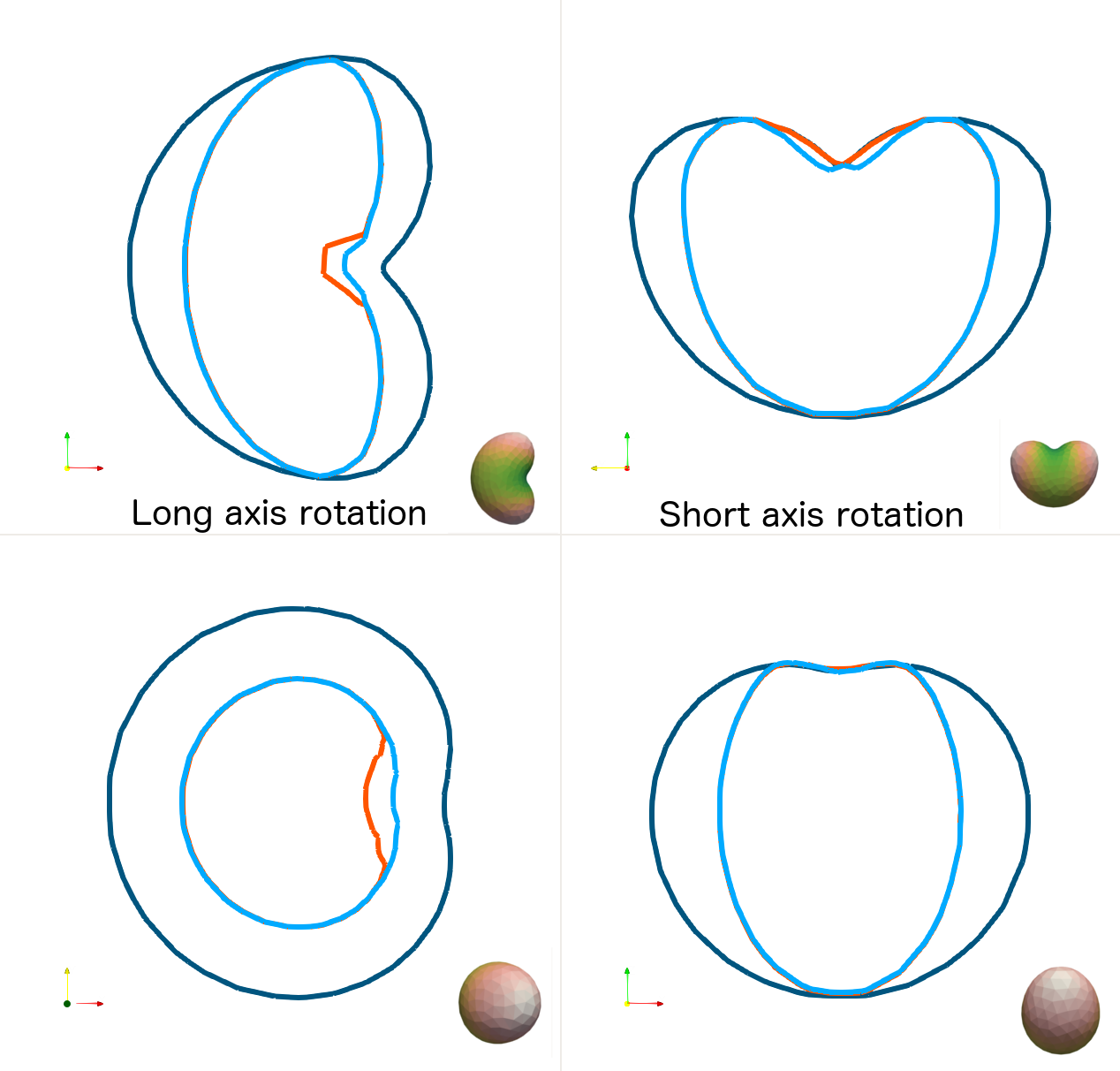}
\caption{Shadowing and self-heating effects on concave shaped objects with different rotation axes. The two panels on the left-hand side show case A) with long axis rotation, with a concave feature on the equator. The two panels on the right-hand side show case B) for short axis rotation with the rotation axis pointing through the center of concave surface structure. The largest contour indicates size and shape of the original object, the smaller (light blue) contour is a resultant shape in case without self-heating, and the red contour is the outline of the final shape with self-heating.}
\label{fig:sdsf.png}
\end{figure}

In Fig.\ref{fig:sdsf.png}, we investigate the effects of shadowing and self-heating on an object with concave feature similar to case (b) in Fig. \ref{fig:concavity.png}, for two different spin axes. The simulation setup is for $\theta=0^{\circ}$, $\varphi=0^{\circ}$, $a=200$~au, and $e=0.3$. The two panels on the left-hand side correspond to case A) with long axis rotation, in which the concavity is located at the equator. The two panels on the right-hand side show case B) with short axis rotation, in which the concave feature is located in the polar region (winter hemisphere). 

 The spin axis is perpendicular to the orbital plane in both cases. All the curves in this plot show an outline of the cross-section of the original body (dark blue), modified shape without self-heating (light blue), and the final shape with self-heating included (red). The red curves clearly illustrate the additional mass loss in the concave region due to self-heating. We can also observe in  Fig.\ref{fig:sdsf.png} the different end results depending on the location of the concavity in case A) versus case B). 
 
 For the long axis case we shown on the left in Fig. \ref{fig:sdsf.png}, the concave structure locates in the equator of the body. Since the obliquity is $0^\circ$ in this case, the sub-sorlar point moves around the equator periodically as the nucleus rotates, the concave center therefore does not experience the shadow (other than nighttime) for most of the time. Meanwhile, as the surrounding region are also illuminated, the self-heating effect plays an important role and enhances mass loss of this region. In the other case, with short axis rotation and a obliquity of 0$^\circ$, the concave center receives no direct illumination, being in shadow, however due to self-heating of the nearby areas the erosion is non-zero. Here, shadowing effect is more significant than self-heating. As a result the erosion process is slower in the concave region in this configuration. 
 In  the scenario when the concave feature locates in the summer pole with non-zero obliquity, the behavior is similar to the bilobed shape (short axis rotation) in  Fig.~\ref{fig:shape_axs.png}. The sublimation process (of homogeneous body) does not completely remove/flatten the concavity.


\subsection{Additional case for a bilobed shape}
\begin{figure}
\includegraphics[width=\columnwidth]{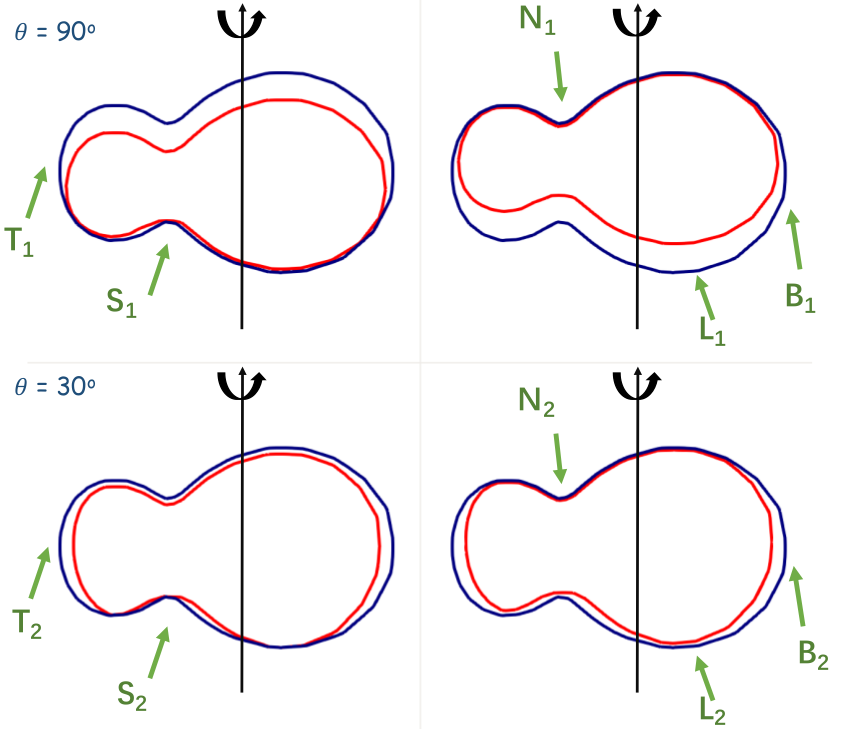}
\caption{Shape evolution for bilobed shape with two lobes of different size, as shown in Fig.~\ref{fig:simulation_shape.png}, with short axis spin. The upper two panels show case for obliquity of 90$^\circ$ and lower ones are for obliquity of 30$^\circ$. The dark blue profile shows the original shape and the red curve shows the evolved one, the rotation axis is marked in black. We shifted the changed shape to align with the original one in order to show the erosion status of the neck for both southern (left panels) and northern part (right panels). The green arrows point to regions where differences with the original shape are notable and different for the two cases of obliquity. They are contrasted in paris (T$_{1}$,T$_{2}$), (N$_{1}$,N$_{2}$), (S$_{1}$,S$_{2}$), (L$_{1}$,L$_{2}$)
 and (B$_{1}$,B$_{2}$), and are discussed in the text. }
\label{fig:bi_asy.png}
\end{figure}

As a last point in our investigations, we present a case of shape modification for a bilobed body in spin along the short axis, however, with different lobe sizes, and with a ``neck'' region offset from the spin axis. The motivation for this run is to see how the ``neck'' location and different sized lobes affect the results we have discussed in connection with Fig.~\ref{fig:shape_axs.png}.  The simulation setup is as follows $\varphi=0^{\circ}$, $a=200$~au, $e=0.3$, for two cases $\theta=90^{\circ}$, and $30^\circ$ as shown in Fig.~\ref{fig:bi_asy.png}. The upper two panels show cases for obliquity of 90$^\circ$ and lower ones are for obliquity of 30$^\circ$. The dark blue line traces an outline of the original shape while the red curve shows the evolved shape. The rotation axis is marked as the arrow-line. In order to better show the results of sublimation driven erosion, we shifted the final shape to align with the original in the south hemisphere (left-hand panels), and northern hemisphere (right-hand side panels). We will discuss interesting features as pointed to by the green arrows that come in pairs (e.g. T$_{1}$, T$_{2}$). T and B indicate the furthest surface region on both side of smaller and larger lobes, respectively. N and S indicate the northern and southern part of the neck region while L is pointing to the southern "belly" region of the larger lobe, which would experience a similar evolution process to that of the smaller lobe.


As expected, for obliquity of 90$^\circ$, the erosion is mainly focused at higher latitude region of southern (summer) hemisphere of the shape, and it shows a larger mass loss compared to the smaller obliquity case in this region (contrasting L$_1$ with L$_2$). In the case of lower obliquity the mass loss rate distribution is more complex, depending on whether the small or big lobe provide shadowing, hence strongly affected by these effects as discussed below.

When we contrast regions pointed to by (green) arrows (T$_{1}$, T$_{2}$ and B$_1$ with B$_2$), we see that the erosion is smaller on both side of the lobes for large obliquity case. However, by comparing the "belly" of the lobes in left panels to right ones, we found that both cases demonstrate the development of north/south dichotomy in both smaller and larger lobes. 

In addition, the ``neck'' region in the southern hemisphere (S$_{1}$, S$_{2}$) is developing differently for the two cases. In the case of $\theta=90^\circ$, the neck region is under illumination with small solar incidence angle for most of the time near perihelion, in which case shadowing effect is unimportant while self-heating effect is more significant. Moreover, by comparing S$_{1}$ with N$_{1}$. we can tell the depth changes on both edges of the neck of the two different lobes are different, which is unlike what happens with bilobed shape with two lobes of the same size for the same rotation states. The erosion process is much more faster on the side of the smaller lobe, this is the result of the self-heating effect. Compared to the larger lobe, the near neck region on the smaller lobe side can ``see'' more facets on the larger lobe and hence gets more re-radiation flux from them. As the illumination on northern hemisphere is weak during perihelion, this effect is not significant in the northern neck region, as N$_{1}$ shows. For $\theta=30^{\circ}$, the erosion in southern neck region (S$_{2}$) also shows asymmetry, but in this case shadowing effect plays a more important role because of the obliquity. The smaller lobe edge experiences a larger degree of shadowing from the bigger lobe, hence, a stronger erosion is in fact experienced by the larger lobe south hemisphere neck.


\section{Discussion}
In this work we set out to investigate the role of illumination driven mass loss on shape change for homogeneous bodies, and various configurations of initial orbital, and spin-axis orientation parameters. 

This study was partly motivated by the recent literature results addressing the bilobed shape of 67P/Churyumov-Gerasimenko visited by the Rosetta spacecraft \citep{Sierks:2015a}, but also the more recent spectacular picture of the KBO object from new Horizons \citep{Stern:2019}. The overarching question is whether solar driven mass loss may alter significantly the overall shape (even at large heliocentric distances), is it possible for a ``neck'' or other concavity be driven purely by sublimation, and does homogeneous mass loss smooth out or deepen pre-existing concavities, and similar questions. In contrast to the recent work of \citet{Vavilov:2019}, we try to approach the solar driven mass loss in a general manner, rather than trying to fit a particular cometary observation. We attempt to explore it as a phenomena valid for any ice, orbits, shapes, etc, and explicitly avoid any assumption on the (ad-hoc) prior localization of activity. 


We considered three basic shapes of homogenous (CO) icy objects, sphere, elongated and oblate, and two more complex shapes, concave and bilobed. Our numerical simulations include both, the shadowing and self-heating effects for shapes with concave structures. By isolating the key orbital and spin-axis parameters, we learned several lessons that should carry over to studies with more advanced thermophysical models. And, although our study used assumption of CO ice driven sublimation, the key characteristics discovered in our numerical experiments are general and should be valid the same for H$_{2}$O (discussed in appendix \ref{sec:appendix}) and CH$_4$ driven mass loss.

\subsection{Summary of results}
What we have learned from our numerical experiments about sublimation driven mass loss can be summarized as follows:
\begin{enumerate}
    \item A north/south hemispheric dichotomy is not an automatic result of illumination driven mass loss for bodies with obliquity and/or large eccentricity. We found that the strongest north/south dichotomy appears when the aphelion of an object is at distances where the mass loss rate function drops super-sharply with decreasing solar flux. In this sense, eccentricity, obliquity are necessary but not sufficient condition.
    \item The integrated mass loss ratio between aphelion to perihelion portion of the orbit is a good indicator that sublimation could be responsible for observed north/south dichotomy (see Fig.~\ref{fig:fluxintegral.png}). Values of this ratio less than 0.5 are sufficient for illumination driven mass loss to create this N/S dichotomy.
    \item The exact pattern of north/south asymmetry is also sensitive to the spin-axis of the body and its orientation with respect to perihelion (longitude of spin axis) as demonstrated in Figs.~\ref{fig:alfdel_sphere.png}, and ~\ref{fig:shape_axs.png}.
    \item We did not find any configuration of orbital and spin axis parameters which would result in the formation of large scale concavity from an initially convex body with homogeneous composition.
    \item Pre-existing concavities are never completely ``erased'' in our simulations, and they can be significantly deepened due to self-heating effects under favourable orientations. In general, the effects of shadowing and self-heating do play a role influencing mass loss distribution even at large heliocentric distances when acting over long time-periods.
    \item The relative sizes of lobes for a bilobed body play a role on the distribution of mass loss rate around its neck region. This effect is strongly coupled to obliquity and spin axis orientation.
    \item A homogeneously active bilobed shape with a narrow (relatively small to the lobe size) ``neck'' should experience separation of the lobes. Again, the role of self-heating for certain spin-orbit orientations will lead to a rapid mass loss.
    \item A sublimation driven mass loss may lead to flattening of initially convex shape segments.
    \item The cases and conclusions presented for sublimation driven mass loss due to CO is applicable for H2O ice, the heliocentric distances where the shape modifications should occur are scaled accordingly (appendix \ref{sec:appendix}).
\end{enumerate}

\subsection{Model approximations and limitations}
The nature of numerical simulations implies accepting a number of simplifying assumptions in order to reduce the number of free parameters in a model. Our aim is to isolate the effects of orbital and spin-axis orientation parameters on the mass loss distribution on analogues of 3D cometary/Kuiper Belt objects due to solar driven sublimation. The question is what kind of shape modification a ``weak'' but long-term sublimation may produce on (non-)spherical bodies, and what kind of end shapes, phenomenologically speaking, can be resulted due to such activity. A study to comprehensively investigate the mass loss distribution of basic 3D shapes due to sublimation does not yet exist.

The perhaps most idealistic simplification is our consideration of pure icy object with zero thermal inertia (although number of studies employ the same for reasons discussed). Although we consider our study to symbolize analogues of Kuiper Belt bodies, we do not yet have reliable knowledge on the physical properties of composition, structure and properties of refractory material of these objects. Same gap in our knowledge pertains to whether the (micro) physical properties can be considered the same across different populations in the Kuiper Belt \citep{Tegler:2008,Brown:2011,Pexinho:2015}. The uncertainties regarding the microphysics of the dust layer, mixing characteristics of ice-dust, surface dust release mechanism, and etc, are very large (e.g. \citep{Skorov:2017}) to justify their use in our study. These complex models may be better employed to study particular object or observations such as from the Rosetta spacecraft. 

Therefore, our results of CO mass loss and shape modification are only linked to the simulation time, ranging from 0.1 to about 100~My years, during which virtually numerical object can loose very significant fraction of its radius. From a dimensional analysis (e.g. \citet{Jewitt:2009}) $\Delta R$~[m] can be estimated as $\approx z/\rho * t$, where $z=$[kgm$^{-2}$s$^{-1}$], $\rho=$[kgm$^{-3}$], and t=[sec], which gives us the very upper limit of $\Delta R \approx$~12~km at 200~au for CO mass loss in 100~My years, in accordance with our simulation time. Nevertheless, qualitatively speaking, results of this study may be extended in validity in the context of including a ``thin'' or ``thick'' dust layer which yields a reduced permeability, and hence mass loss rate. For instance, if such a layer reduces gas mass loss by a factor of 5-10, under the assumption that the nucleus' composition-structure and the thickness of the dust layer do not change during evolution, the simulation time can be extended by the same amount to reach the same mass loss, although this scaling has its obvious limits of validity. This approach is technically similar to the models B, C in \citet{Keller:2015}. The reduced permeability for these models will not influence our conclusion on the shape modifications structure, they would only imply longer times to reach the same mass loss. 

Second limitation to be mentioned is related mostly to the self-consistency of the spin-axis coupling to orbital and sublimation induced torques. Most of the difficulty is technical, and rests on the need to take into account two time scales that are very different, but clearly coupled. The time scale of orbital period is about hundreds to millions of years in our simulations, while the rotation period of these bodies is on the order of hours (10 hours or so). In order to calculate the sublimation torque and solve the Euler equation, a very small time step is required to achieve an accurate integral of the variations of the orientation of spin axis and spin rate. 

\citet{scheeres:2007} and \citet{nesvorny:2007} established averaging methods for YORP effects experienced by small bodies (mostly asteroids) taking real time insolation and shape into account. However, unlike thermal radiation, as Fig~ \ref{fig:fluxdist.png} shows, the CO (or water) production rate variation versus solar distance is too complex to be written into a linear or an polynomial formula (over the entire range we want to investigate). \citet{sidorenko:2008} applied the averaging method to obtain evolutionary equations in order to study the long-term variations in the nucleus spin state induced by outgassing torques, but the illumination variation is not considered in their simplified model. There are further analytic approaches described in literature \citep{samarasinha:2013, steckloff:2016, steckloff:2018}, however, these models predicted nucleus' spin state change without taking the exact 3D shape model and its variation induced by mass loss into consideration. Therefore, to our knowledge, there exists no approximate framework which is able to combine spin, mass loss together with orbital evolution for the 3D shapes (accounting for shadowing and self-heating) we are exploring in this paper. Currently, only fully numerical schemes are appropriate to investigate sublimation torques in our context. At the same time, it is not feasible to include such numerically expensive consideration when exploring the parameter space of orbits, mass loss function, and axis orientations. Therefore, in this work, we consider nucleus' rotation only for the stable case, without taking the wobbling rotation excitation and spin states change due to sublimation torque into consideration. However, we plan to perform these in the future, relying on the knowledge gained on physically and scientifically interesting cases from this work. 
As a general remark, the sublimation torque induced spin (and/or orientation) changes are expected not to play a role only if the orientation variation time scale is significantly larger than the time scale of shape modification. On the other hand, if the time scale is significantly shorter, new results are expected since the variation period of $\theta$ and $\varphi$ might bring new periodical effects. A tumbling spin state with large procession angle or even a chaotic rotation may average the mass loss on the surface to yield an ``averaged'' shape as a result. Therefore, if a nucleus exhibits an asymmetry shape believed to be driven by sublimation activity, it would indicate that the nucleus rotates in stable stage for a long term period in its dynamical history. In other words, this implies that the time scale of wobbling excitation of such objects might be much longer than the one for shape modification.

\subsection{Future work}
The future work should focus on two important points discussed in the text. First, to include sublimation torques in a self-consistent way into the existing code. Second,  to include a more detailed (physical) thermophyiscal model to study morphological changes, for example, for Centaurs and/or JFCs. Furthermore, including a re-meshing routine into the code, might be important especially in detailed study on the role of mass loss on the evolution of concavities and/or pits on the surface (different depths, sizes and slopes). 


\section*{Acknowledgements}
We would like to thank H. Keller for his encouragement and many discussions on the several topics presented here. YZ thanks Dr. L. Yu for many helpful discussions. YZ is supported from the National Natural Science Foundation of China (Grant No: 11761131008, 11673072, 11633009), the Strategic Priority Research Program on Space Science(CAS) (Grant No. XDA15017600) and the Foundation of Minor Planets of Purple Mountain Observatory. LR was supported from the project DFG-392267849. We thank the two anonymous reviewers for their constructive comments and careful reading of the manuscript. The suggested revisions and additions made the paper clearer and enhanced the overall quality of the paper.




\bibliographystyle{mnras}
\bibliography{jfc1.bib} 



\appendix
\section{Effects of eccentricity for water ice}
\label{sec:appendix}
\begin{figure}
\includegraphics[width=\columnwidth]{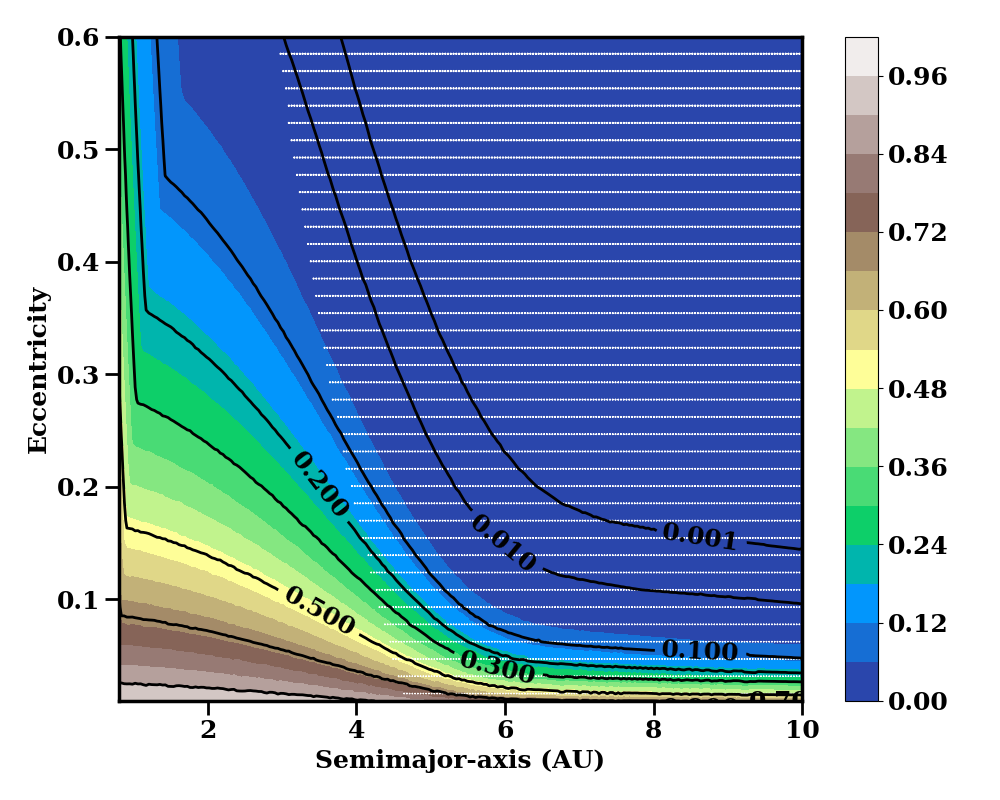}
\caption{A ratio of H$_{2}$O mass loss rate at apogee to perigee (colorbar) as a function of eccentricity and semi-major axis. The overlaid white dots denote a region in which aphelion distance is beyond the ``knee'' in the water mass loss rate function (taken to be at 4.7~au). See text for discussion.}
\label{fig:fluxratioh2o.png}
\end{figure}

Below we present an analysis of the key orbital parameters $(a,e)$ in combination with the mass loss rate function for pure H$_{2}$O ice in order to draw parallel with the figures ~\ref{fig:fluxratio.png},~\ref{fig:fluxintegral.png} done for CO ice (in the main text). When considering a different kind of ice (as shown below), the heliocentric distance where a strong north/south dichotomy may develop due to sublimation driven activity occurs at other configuration in the $(a,e)$ space. The Fig.~\ref{fig:fluxratioh2o.png} highlights this very point.  

\begin{figure}
\includegraphics[width=\columnwidth]{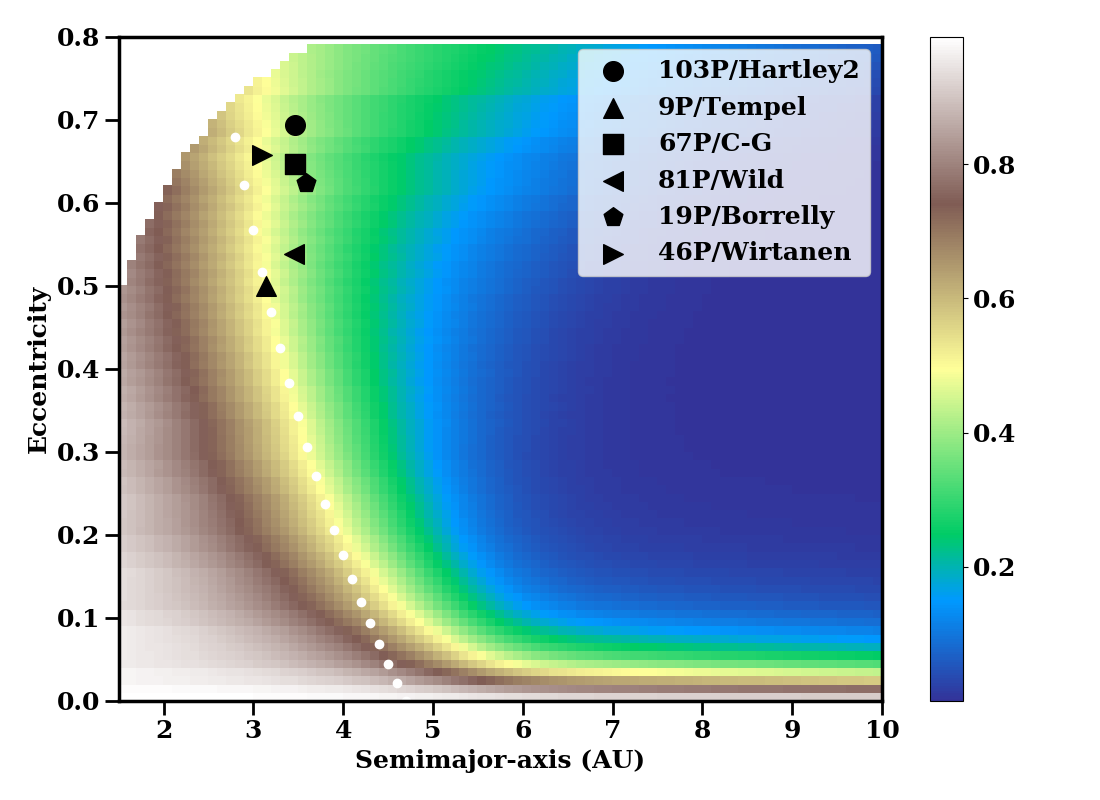}
\caption{A ratio of integrated H$_{2}$O mass loss for aphelion to perihelion segment of the orbit, as a function of eccentricity and semi-major axis. This figure corroborates Fig.~\ref{fig:fluxratioh2o.png}, and provides another view for what configurations of ($a,e$) can a strong hemispherical dichotomy occur. The white dotted line indicates a boundary (in geometric sense) beyond which the aphelion cross the ``knee'' of the H$_{2}$O mass loss rate function. The black symbols show several well known Jupiter-Family comets. See text for discussion. }
\label{fig:fluxintegralh2o.png}
\end{figure}

A more quantitative way of looking at this is to compute a ratio of integrated mass loss for $90^{\circ} < f \le 270^{\circ}$, when the sub-solar point is mostly (depends on orientation) on the winter hemisphere and $f \le 90^{\circ}$ or $f > 270^{\circ}$ when the sub-solar point is mostly on the summer hemisphere,
which is shown in Fig.~\ref{fig:fluxintegralh2o.png}. As discussed in the case of CO, the similar conclusion applies for H$_{2}$O. Specifically, in the $(a,e)$ space where this ratio of integrated mass loss is below 0.5, the conditions are such that the bodies may develop a strong hemispherical dichotomy due to sublimation activity. We overplot several well known comets in this figure, including that of 67P/C-G visited by the Rosetta spacecraft, which is known to have a strong north-south morphological dichotomy. From this point of view comets 67P/C-G, 19P/Borrelly, 103P/Hartley 2 all share a similar (and relatively low) value of the ratio. For example, comet 9P/Tempel, which is known to have spherical type of shape with small hemispherical dichotomy indeed lies just at the border with a ratio of about 0.5. In fact, most of the JFCs are expected to lie close to this border by the nature of their orbit, hence, suitable conditions exists for north/south asymmetry due to a water sublimation mass loss. However, we should keep in mind that our numerical results apply only to homogeneous bodies with surface ice (and zero thermal inertia). The implicit assumption is also that such bodies stay on the same orbit for long enough time for the north/south dichotomy to appear, without changes to the spin-axis orientation.


\section{Validation of the shape modification method}
\subsection{Inverse propagation}
\label{sec:appendix_va1}

\begin{figure}
\includegraphics[width=\columnwidth]{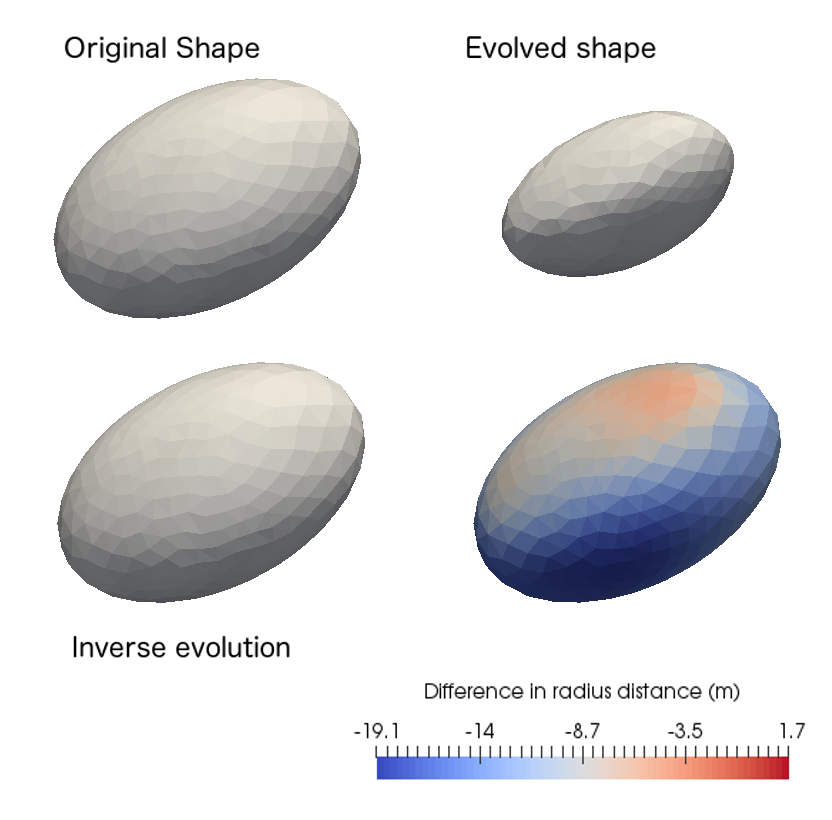}
\caption{A comparison between original and backward propagated shapes. The upper left panel shows the original shape. In the upper right panel one can see a result after 12,000 orbits, while in the lower left panel a final shape is presented after such body is propagated backwards for 12,000 orbits. The difference in radius distance of surface vertices between the original (upper left) and the back-evolved shape (lower left) is shown in the lower right panel. The differences are smaller than 0.5\%. The color bar denotes the differences in absolute scale [metres].}
\label{fig:compare_reverse.png}
\end{figure}

The first approach of validating our shape modification implementation was by comparing an original and backward-evolved shape. The result and description is  presented here for an elongated shape in Fig~\ref{fig:compare_reverse.png}.

\subsection{Comparison with analytical solution}
\label{sec:appendix_va2}

\begin{figure}
\includegraphics[width=\columnwidth]{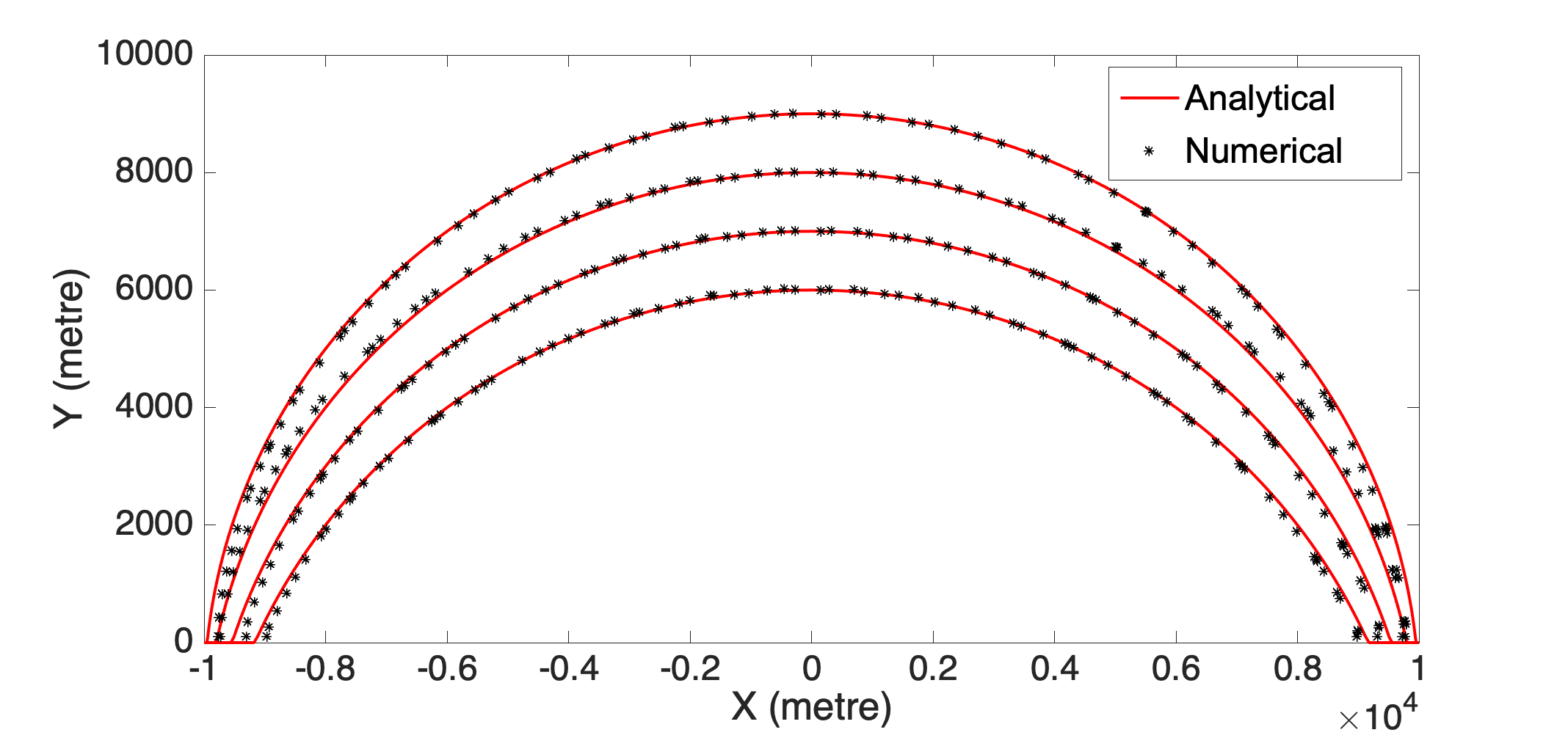}
\caption{Outlines of a shape evolution for a spherical body comparing numerical method used in this paper and an analytical method adopted in a simplified form from \citet{Vavilov:2019}.  Red curves are results derived from analytical solution and black stars are data extracted from evolved shape using numerical method. The mass loss rate in equator is 0.1~m per orbit and the curves from top to bottom show results after 10,000, 20,000, 30,000, 40,000 orbits, respectively.}
\label{fig:compare_analytical.png}
\end{figure}

Another approach to validate our shape modification procedure is to compare the numerical and analytical solutions for shape change of a convex, rotational symmetric body. In the following scenario, we assume a spherical body with its spin axis perpendicular to the orbital plane. The coordinate system is defined as follows: the $\widetilde{x}$~axis is s oriented along the sphere's rotation axis while the $\widetilde{y}$~axis points in the direction of perihelion. In the cross section of the nucleus on the $x-y$ plane  (Fig.~\ref{fig:compare_analytical.png}), $y(x,t)$ indicates surface distance from the spin axis, where t denotes the evolution time. Furthermore, for a rotational symmetric body, the change of the function y(t,x) describes the change of the shape.

To simplify the validation, we neglect the nonlinear change of mass loss curve (Fig~\ref{fig:fluxdist.png}) and avoid the calculation of average mass loss rate as presented, for example, in the appendix of \cite{Vavilov:2019}. Instead we follow a linear approach by defining averaged depth change rate as
\begin{equation}
Z(x,y)=Z_0\cos\theta,
\end{equation}
where $\theta$ is the solar incidence angle and $Z_0$ is the depth change rate on the equator averaged over one orbit. Under these assumptions and following Eq.A1 in appendix of \cite{Vavilov:2019}, we have the solution for previous function in the form of
\begin{equation}
y(t,x) = y(0,x) - Z_0t.
\end{equation}
In both of our analytical and numerical calculations, $Z_0$ has a value of 0.1 metre per orbit.

The results derived by both methods are presented in Fig.~\ref{fig:compare_analytical.png}. The profiles from top to bottom describe shape evolution in the cross section of $x-y$ plane after 10,000, 20,000, 30,000, 40,000 orbits, respectively. The red lines correspond to the analytical solution and black stars are the outlines extracted from our numerical results for a 3D shape. The results show good agreement which provides further justification for our shape change implementation.

\bsp	
\label{lastpage}

\end{document}